\documentstyle[12pt]{article}

\begin{document}
\title{Coefficients of Chiral Perturbation Theory}
\author{Bing An Li\\
Department of Physics and Astronomy, University of Kentucky\\
Lexington, KY 40506, USA}

\maketitle

\begin{abstract}
Based on an effective chiral theory of pseudoscalar, vector, and
axial-vector mesons, the coefficients of the chiral perturbation
theory are predicted. There is no new parameter in these predictions.
The current
quark masses, $m_{\eta}$, $f_{K}$, and $f_{\eta}$ are determined
too.
\end{abstract}

\newpage
The chiral perturbation theory(ChPT)
proposed a decade ago[1] is successful
in parametrizing low-energy $QCD$. The chiral symmetry revealed from
$QCD$, quark mass expansion, and momentum expansion
are used to construct the Lagrangian of ChPT
\begin{eqnarray}
\lefteqn{{\cal L}={f^{2}_{\pi}\over16}TrD_{\mu}UD^{\mu}U^{\dag}+
{f^{2}_{\pi}\over16}Tr\chi(U+U^{\dag})+
L_{1}[Tr(D_{\mu}UD^{\mu}U^{\dag})]^{2}+L_{2}(TrD_{\mu}UD_{\nu}U^
{\dag})^{2}}\nonumber \\
&&+L_{3}Tr(D_{\mu}UD^{\mu}U^{\dag})^{2}+L_{4}Tr(D_{\mu}U
D^{\mu}U^{\dag})Tr\chi(U+U^{\dag})
+L_{5}TrD_{\mu}UD^{\mu}U^{\dag}(\chi U^{\dag}+U\chi)\nonumber \\
&&+L_{6}[Tr\chi(U+U^{\dag})]
^{2}+L_{7}[Tr\chi(U-U^{\dag})]^{2}+L_{8}
Tr(\chi U\chi U+\chi U^{\dag}\chi U^{\dag})
\nonumber \\
&&-iL_{9}Tr(F^{L}_{\mu\nu}D^{\mu}UD^{\nu}U^{\dag}+F^{R}_{\mu\nu}
D^{\mu}U^{\dag}D^{\nu}U)+L_{10}Tr(F^{L}_{\mu\nu}UF^{\mu\nu R}U^{\dag}).
\end{eqnarray}
The parameters
in the chiral perturbation theory(1) are pion decay constant
$f_{\pi}$ and the 10 coefficients. These parameters are determined
by fitting experimental data[1,2,3,4,5,6,7].
\begin{table}[h]
\begin{center}
\caption{Values of the coefficients[1(b)]}
\begin{tabular}{|c|c|c|c|c|c|c|c|c|c|} \hline
$10^{3}L_{1}$&$10^{3}L_{2}$&$10^{3}L_{3}$
&$10^{3}L_{4}$&$10^{3}L_{5}$&$10^{3}L_{6}$
&$10^{3}L_{7}$&$10^{3}L_{8}$&$10^{3}L_{9}$
&$10^{3}L_{10}$  \\ \hline
$.9\pm.3$&$1.7\pm.7$&-$4.4\pm 2.5$&$0.\pm.5$&
$2.2\pm.5$&$0.\pm0.3$&-$.4\pm.15$& $1.1\pm.3$&
$7.4\pm.7$&-$6.\pm.7$ \\ \hline
\end{tabular}
\end{center}
\end{table}

The chiral perturbation theory is rigorous and phenomenologically
successful in describing the physics of the pseudoscalar mesons at low
energies. Models attempt to deal with the two main frustrations that
the ChPT is limited to pseudoscalar mesons at
low energy and contains many coupling
constants which must be measured. The price paid for including more
mesons and determining the coefficients is
to make additional assumption. Many models try to
predict the chiral couplings(see Table 2[8]). As pointed in Ref.[8],
the most reliable one
appears to be the influence of the low-lying
resonances[4]. There are many attempts to construct an effective
theory containing vector and axial-vector mesons too[9]. Obviously,
a successful extension should recover the results obtained by the ChPT
and predict the coefficients of the ChPT.

\begin{table}[h]
\begin{center}
\caption{Coefficients obtained by models}
\begin{tabular}{|c|c|c|c|c|c|c|c|} \hline
&  Expt&Vectors[4]&Quark[2]&Ref.[6]&Ref.[7]&Nucleon loop&Linear$\sigma$
model \\ \hline
$L_{1}$+${1\over2}L_{3}$&-.9&-2.1&-.8&2.1&1.1&-.8&-.5  \\ \hline
$L_{2}$ & 1.9&2.1&1.6&1.6&1.8&.8&1.5  \\ \hline
$L_{9}$ & 7.1&7.3&6.3&6.7&6.1&3.3&.9 \\ \hline
$L_{10}$ & -5.6&-5.8&-3.2&-5.8&-5.2&-1.7&-2.0  \\ \hline
\end{tabular}
\end{center}
\end{table}

I have proposed an effective chiral theory of pseudoscalar, vector, and
axial-vector mesons[10,11]. It provides a unified study of
meson physics at low energies. It is used
to predict all the coefficients of the ChPT in this paper.
Before doing that it
is meaningful to present more detailed arguments about
the ansatz made in this theory. The ansatz is that
the meson fields are
simulated by quark operators.
For example,
\begin{equation}
\rho^{i}_{\mu}=-\frac{1}{g_{\rho}m^{2}_{\rho}}\bar{\psi}\tau_{i}
\gamma_{\mu}\psi.
\end{equation}
The ansatz(2) can be tested. Applying
PCAC, current algebra, and this expression(2) to the
decay $\rho\rightarrow\pi\pi$, under the soft pion approximation
it is derived
\begin{equation}
{1\over 2}f_{\rho\pi\pi}g_{\rho}=1
\end{equation}
which is just the result of the VMD[12]. As a matter of fact, using
fermion operator to simulate a meson field has a long history.
More than six decades ago, Jordan et al[13] observed
\begin{equation}
{1\over \sqrt{\pi}}\partial_{\mu}\phi=\bar{\psi}\gamma_{5}\gamma_{\mu}
\psi
\end{equation}
in $1+1$ field theory. A similar relation as Eq.(4) is proved in the
bosonization of 1+1 field theory[14].
On the other hand, using quark operators to simulate meson fields
have been already exploited in the Nambu-Jona-Lasinio(NJL) model[15],
a model of four quark interactions.

In Refs.[10,11]
the simulations of the meson fields by quark operators are realized by
a Lagrangian which
is constructed by
using chiral symmetry and the minimum
coupling principle
\begin{eqnarray}
{\cal L}=\bar{\psi}(x)(i\gamma\cdot\partial+\gamma\cdot v
+\gamma\cdot a\gamma_{5}
-mu(x))\psi(x)-\bar{\psi(x)}M\psi(x)\nonumber \\
+{1\over 2}m^{2}_{0}(\rho^{\mu}_{i}\rho_{\mu i}+K^{*\mu}K^{*}_{\mu}+
\omega^{\mu}\omega_{\mu}+\phi^{\mu}\phi_{\mu}
+a^{\mu}_{i}a_{\mu i}+K^{\mu}_{1}K_{1\mu}
+f^{\mu}f_{\mu}+f^{\mu}_{1s}f_{1s\mu})
\end{eqnarray}
where M is the quark mass matrix
\[\left(\begin{array}{c}
         m_{u}\hspace{0.5cm}0\hspace{0.5cm}0\\
         0\hspace{0.5cm}m_{d}\hspace{0.5cm}0\\
         0\hspace{0.5cm}0\hspace{0.5cm}m_{s}
        \end{array}  \right ),\]
\(v_{\mu}=\tau_{i}\rho^{i}_{\mu}+\lambda_{a}K^{*a}_{\mu}
+({2\over3}+{1\over\sqrt{3}}\lambda_{8})\omega_{\mu}
+({1\over3}-{1\over\sqrt{3}}\lambda_{8})\phi_{\mu}\),
\(a_{\mu}=\tau_{i}a^{i}_{\mu}+\lambda_{a}K^{a}_{1\mu}
+({2\over3}+{1\over\sqrt{3}}\lambda_{8})f_{\mu}+({1\over3}
-{1\over\sqrt{3}}\lambda_{8})f_{1s\mu}\),
and \(u=exp\{i\gamma_{5}(\tau_{i}\pi_{i}+\lambda_{a}K_{a}+\lambda_{8}
\eta_{8}+\eta_{0})\}\).
$u$ can be written as
\begin{equation}
u={1\over 2}(1+\gamma_{5})U+{1\over 2}(1+\gamma_{5})U^{\dag},
\end{equation}
where \(U=exp\{i(\tau_{i}\pi_{i}+\lambda_{a}K_{a}+\lambda_{8}
\eta+\eta_{0})\}\).
Since pions, kaons
and $\eta$ are Goldstone bosons we treat them differently
from vector and axial-vector mesons. Through the scheme of the
nonlinear $\sigma$ model the pseudoscalar mesons are
introduced to the Lagrangian(5).
This treatment is backed by the ChPT.
In the Lagrangian of the ChPT(1) the first two terms
are from the nonlinear $\sigma$ model.
The matrix representations of the pseudoscalar mesons,
$U$ and $U^{\dag}$, in Eq.(5) are the same as the ones in the ChPT(1).
Spontaneous chiral symmetry
breaking is revealed from this mechanism[10] when taking \(u\rightarrow
1\).
On the other hand, mesons are bound states solutions of $QCD$ they
are not independent degrees of freedom. Therefore,
there are no kinetic terms for meson fields. The kinetic terms
of meson fields are generated from quark loops[10,11].
Using the least action principle, the relationship between meson fields
and quark operators are found from the Lagrangian(5). Taking the case of
two flavors as an example, from the least action principle
\begin{eqnarray}
\frac{\delta{\cal L}}{\delta\Pi_{i}}=0,\;\;\;
\frac{\delta{\cal L}}{\delta\eta}=0,\;\;\;
\frac{\delta{\cal L}}{\delta\rho^{i}_{\mu}}=0,\;\;\;
\frac{\delta{\cal L}}{\delta a^{i}_{\mu}}=0,\;\;\;
\frac{\delta{\cal L}}{\delta\omega_{\mu}}=0,\;\;\;
\frac{\delta{\cal L}}{\delta f_{\mu}}=0,
\end{eqnarray}
we obtain
\begin{eqnarray}
{\Pi_{i}\over\sigma}=i(\bar{\psi}
\tau_{i}\gamma_{5}\psi+ix\bar{\psi}\tau_{i}\psi)/(\bar{\psi}
\psi+ix\bar{\psi}\gamma_{5}\psi),\nonumber \\
x=(i\bar{\psi}\gamma_{5}\psi-{\Pi_{i}\over\sigma}
\bar{\psi}\tau_{i}\psi)/
(\bar{\psi}\psi+i{\Pi_{i}\over\sigma}
\bar{\psi}\tau_{i}\gamma_{5}
\psi),\nonumber \\
\rho^{i}_{\mu}=-{1\over m^{2}_{0}}\bar{\psi}\tau_{i}\gamma_{\mu}
\psi,\;\;\;
a^{i}_{\mu}=-{1\over m^{2}_{0}}\bar{\psi}\tau_{i}\gamma_{\mu}
\gamma_{5}\psi,\nonumber \\
\omega_{\mu}=-{1\over m^{2}_{0}}\bar{\psi}\gamma_{\mu}
\psi,\;\;\;
f{\mu}=-{1\over m^{2}_{0}}\bar{\psi}\gamma_{\mu}
\gamma_{5}\psi,
\end{eqnarray}
where \(\sigma+i\gamma_{5}\tau\cdot\Pi=
ue^{-i\eta\gamma_{5}}\)
, \(\sigma=\sqrt{1-\Pi^{2}}\), and \(x=tan\eta\).
The pseudoscalar fields have very complicated quark structures.
Substituting the expressions(8) into the Lagrangian(5) of two flavors,
a Lagrangian of quarks is obtained. It is no longer a theory of four
quark interactions. The vector and axial-vector parts
are the same as the ones in the NJL model. However,
the pseudoscalar part is different from the NJL model.

As shown in Refs.[10,11] there are explicit chiral symmetry(PCAC[19]
), dynamical chiral symmetry breaking, and large
$N_{C}$ expansion. They originate in $QCD$. However,
theoretically,
we do not know how to prove the ansatz of simulating
the meson fields by quark operators for $1+3$ $QCD$.
We use the Lagrangian(5)
to study meson physics at low
energies to see whether the Lagrangian is phenomenologically
successful. In Refs.[10,11] the Lagrangian(5) has been used to calculate
the masses and electromagnetic, weak, and strong properties of the
pseudoscalar, vector, and axial-vector mesons. Theoretical results
agree well with the data. Taking the masses of the $\rho$ and the
$a_{1}$ meson
as an example of the results obtained by this theory.
The mass of the
$\rho$ meson is originated in dynamical chiral symmetry breaking[16].
The mass difference of the $\rho$ and the $a_{1}$ mesons is
from spontaneous chiral symmetry breaking[10].

Large $N_{C}$ expansion is part of this effective chiral theory[10].
All
the calculations are done at the tree level and meson loops are at
higher order in large $N_{C}$ expansion[10].
All the masses of the mesons
are less than the cut-off[10].
Therefore, this theory is self-consistent.

By using the ansatz, this effective chiral theory extends the ChPT to
include vector and axial-vector mesons. However, this theory faces the
second challenge: predicting all the coefficients of the ChPT. This is
the task of this paper. The study done in this paper is at the tree
level.

As a matter of fact, in Ref.[10] two of the
coefficients have been determined from the study of $\pi\pi$ scattering
\begin{eqnarray}
\alpha_{2}=-\alpha_{1}=c^{2}, \\
c={f^{2}_{\pi}\over2gm^{2}_{\rho}},
\end{eqnarray}
where $\alpha_{1,2}$ are defined in Ref.[4]
and g is the universal coupling constant of this theory[10,11].
Comparing with Eq.(1),
we obtain
\begin{equation}
\alpha_{1}=4L_{1}+2L_{3},\;\;\;\alpha_{2}=4L_{2}.
\end{equation}
Eq.(9) is the same as the one obtained in Ref.[4] in which the $\rho$
resonance participates in $\pi\pi$ scattering.
From Eq.(11) it is
obtained
\begin{eqnarray}
2(L_{1}+L_{2})+L_{3}=0,\\
L_{2}={1\over4}c^{2}.
\end{eqnarray}

The decay $K_{l4}$ has been used to determine the coefficients of
ChPT[17,18].
Both vector and axial-vector currents contribute to $K_{l4}$.
In this paper we only try to predict $L_{3}$ from $K_{l4}$.
Complete study of $K_{l4}$ is beyond the
scope of this paper.
In Ref.[19] we have presented a bosonized axial-vector current(Eq.(72)
of
Ref.[19]) which has been used to study many $\tau$ mesonic decays.
Theoretical results agree well with data.
Using the axial-vector current
and the vertices $K_{A}K^{*}\pi$, $K^{*}K
\pi$, $K_{A}\rho k$, $\rho K\bar{K}$, and $\rho\pi\pi$ presented in Ref.
[19] in the chiral limit, the form factors of the
matrix element of the axial-vector current of $K_{l4}$
\begin{equation}
<\pi^{+}(p_{+})\pi^{-}(p_{-})|A_{\mu}|K^{+}(k)>=-{i\over m_{K}}\{F(p_{+}
+p_{-})_{\mu}+G(p_{+}-p_{-})_{\mu}+R(k-p_{+}-p_{-})_{\mu}\}
\end{equation}
can be calculated. The calculation is similar to
$\tau\rightarrow K^{*}\pi\nu, K\rho\nu$ done in Ref.[19].
Using Eqs.(40,43) of the form factor G
of Ref.[18], it is derived
\begin{equation}
L_{3}=-{1\over4}\{gc-{3\over2}c^{2}
+{3\over8\pi^{2}}(1-{2c\over g})^{2}\}.
\end{equation}
The same $L_{3}$ is obtained by calculating related term of
the form factor
$F^{+-}_{1}$ defined in Ref.[17].

From Eqs.(12) the $L_{1}$ is obtained
\begin{equation}
L_{1}={1\over8}\{gc-{7\over2}c^{2}
+{3\over8\pi^{2}}(1-{2c\over g})^{2}\}.
\end{equation}
It is learned from Ref.[10] that
$g^{2}\sim O(N_{C})$, $f^{2}_{\pi}\sim
O(N_{C})$, and $m^{2}_{\rho}\sim O(1)$ in large $N_{C}$ expansion.
Therefore, $L_{1,2,3}\sim O(N_{C})$.
The predictions of $L_{1-3}$ are made at the tree level, therefore,
the Eq.(12) actually means that $2(L_{1}+L_{2})+L_{3}\sim O(1)$ in
large $N_{C}$ expansion.

According to Ref.1(b), the coefficients $L_{4-8}$ are determined from
the quark mass expansions of
$m^{2}_{\pi}$, $m^{2}_{K}$, $m^{2}_{\eta}$, $f_{\pi}$, $f_{K}$, and
$f_{\eta}$.
Therefore, the quark mass term of the Lagrangian(5) is needed to be
taken into account in
deriving the effective Lagrangian of mesons.
The
expressions of the masses and decay constants of the pseudoscalars
are found from the real part of the effective
Lagrangian. Using the method presented in Ref.[10],
in Euclidean space the real part of the effective Lagrangian of mesons
with
quark masses is written as
\begin{eqnarray}
\lefteqn{{\cal L}_{RE}={1\over2}\int d^{D}x\frac{d^{D}p}{(2\pi)^{D}}
\sum^{\infty}_{n=1}{1\over n}\frac{1}{(p^{2}+m^{2})^{n}}}
\nonumber \\
&&Tr\{(\gamma\cdot\partial-i\gamma\cdot v+i\gamma\cdot a\gamma_{5})
(\gamma\cdot\partial-i\gamma\cdot v-i\gamma\cdot a\gamma_{5})
+2ip\cdot(\partial-iv-ia)\nonumber \\
&&+m\gamma\cdot Du
-i[\gamma\cdot v,M]+i\{\gamma\cdot a,M\}\gamma_{5}
-m(\hat{u}M+Mu)-M^{2}\}^{n}.
\end{eqnarray}
where \(D_{\mu}u=\partial_{\mu}u-i[v_{\mu},u]+i\{a_{\mu},u\}
\gamma_{5}\) and \(\hat{u}=
exp\{-i\gamma_{5}[\tau_{i}\pi_{i}+\lambda_{a}
K_{a}+\lambda_{8}\eta_{8}+\eta_{0}]\}\).
Comparing with Eq.(11) of Ref.[10],
there are new terms in which the
quark mass matrix M is involved.
As done in Ref.[10], the mesons fields in Eq.(17) are needed to be
normalized.

The masses of the pseudoscalcar mesons are found from the
coefficients of the terms $\pi_{i}\pi_{i}$, $K_{a}K_{a}$, $\eta_{8}
\eta_{8}$ in the effective Lgarngian(17).
The masses of the pseudoscalars
can be calculated
to any order in quark masses.
Up to the second order in quark masses we obtain
\begin{eqnarray}
\lefteqn{m^{2}_{\pi^{\pm}}={4\over f^{2}_{\pi^{\pm}}}\{-{1\over3}
<\bar{\psi}\psi>(m_{u}+m_{d})-{F^{2}\over4}(m_{u}+m_{d})^{2}
\},}\nonumber \\
&&m^{2}_{\pi^{0}}={4\over f^{2}_{\pi^{0}}}\{-{1\over3}
<\bar{\psi}\psi>(m_{u}+m_{d})-{F^{2}\over2}(m^{2}_{u}+m^{2}_{d})
\},\nonumber \\
&&m^{2}_{K^{+}}={4\over f^{2}_{K^{+}}}\{-{1\over3}
<\bar{\psi}\psi>(m_{u}+m_{s})-{F^{2}\over4}(m_{u}+m_{s})^{2}
\},\nonumber \\
&&m^{2}_{K^{0}}={4\over f^{2}_{K^{0}}}\{-{1\over3}
<\bar{\psi}\psi>(m_{d}+m_{s})-{F^{2}\over4}(m_{d}+m_{s})^{2}
\},\nonumber \\
&&m^{2}_{\eta_{8}}={4\over f^{2}_{\eta_{8}}}\{-{1\over3}
<\bar{\psi}\psi>{1\over3}(m_{u}+m_{d}+4m_{s})
-{F^{2}\over6}(m^{2}_{u}+m^{2}_{d}+4m^{2}_{s})\},
\end{eqnarray}
where $F^{2}$ is defined in Ref.[10] as
\[F^{2}={N_{C}\over \pi^{4}}m^{2}\int\frac{d^{4}p}
{(p^{2}+m^{2})^{2}},\]
$<\bar{\psi}\psi>$ is the quark condensate of three flavors.
In terms of a cut-off $\Lambda$ the quark condensate
is expressed as(using Eqs.(40) and (43) of Ref.[10])
\begin{equation}
<\bar{\psi}\psi>={i\over(2\pi)^{4}}
Tr\int d^{4}p\frac{\gamma\cdot p-mu}{p^{2}
-m^{2}}
=\frac{m^{3}N_{C}}{4\pi^{2}}
\{{\Lambda^{2}\over m^{2}}-log({\Lambda^{2}\over m^{2}}+1)\}.
\end{equation}
The universal coupling constant g is expressed as[10]
\begin{equation}
g^{2}={F^{2}\over6m^{2}}={1\over2\pi^{2}}\{log({\Lambda^{2}\over m^{2}}
+1)+\frac{1}{{\Lambda^{2}\over m^{2}}+1}-1\}.
\end{equation}
The decay constants of the pseudoscalars $f_{\pi}$, $f_{K}$, and
$f_{\eta}$ are defined by normalizing the
pseudoscalar fields as done in Refs.[10,11].
In this paper
we calculate the contributions of the quark masses to these
decay constants to $O(m_{q})$.
Using the
effective Lagrangian(17), the decay constants can be calculated
to any order
in quark masses. As indicated in Refs.[10,11],
there is mixing between the
axial-vector field and corresponding pseudoscalar field. The mixing
results in the shifting of the axial-vector field $a_{\mu}$
\begin{equation}
a_{\mu}\rightarrow {1\over g_{a}}a_{\mu}
-{c_{a}\over g_{a}}\partial_{\mu}P,
\end{equation}
where P is the corresponding pseudoscalar field, $g_{a}$ is the
normalization constant of the $a_{\mu}$ field, and $c_{a}$
is the mixing
coefficient. Both $g_{a}$ and $c_{a}$ are determined
in the chiral limit in Ref.[10].
For pion and $a_{1}$ fields
it is obtained by eliminating the mixing between pion
field and $a_{1}$ field
\begin{equation}
{c_{a}\over g_{a}}={1\over g^{2}_{a}m^{2}_{a}}\{{F^{2}\over2}+
({F^{2}\over8m^{2}}-{3\over4\pi^{2}})2m(m_{u}+m_{d})\},
\end{equation}
where $m_{a}$ is the mass of the $a_{1}$ meson which is determined from
the Lagrangian(17) as
\begin{equation}
g^{2}_{a}m^{2}_{a}=F^{2}+g^{2}m^{2}_{\rho}+6g^{2}_{0a}m(m_{u}+m_{d}),
\end{equation}
where $g^{2}_{0a}$ is expressed as[10]
\begin{equation}
g^{2}_{0a}=g^{2}(1-{1\over2\pi^{2}g^{2}}).
\end{equation}
Up to the first order in quark masses,
the decay constant $f^{2}_{\pi}$ is obtained from the Lagrangian(17)
\begin{equation}
f^{2}_{\pi}=f^{2}_{\pi0}\{1+f\frac{m_{u}+m_{d}}{m}\},
\end{equation}
where
\begin{equation}
f^{2}_{\pi0}=F^{2}(1-{2c\over g})[10]
\end{equation}
and
\begin{equation}
f=(1-{2c\over g})(1-
{1\over2\pi^{2}g^{2}})-1+\frac{4}{\pi^{2}f^{4}_{\pi0}}(-{1\over3})
<\bar{\psi}\psi>m(1-{2c\over g})(1-{c\over g}),
\end{equation}
In the same way, $f^{2}_{K^{+}}$, $f^{2}_{K^{0}}$, and
$f^{2}_{\eta_{8}}$ are found
\begin{eqnarray}
f^{2}_{K^{+}}=f^{2}_{\pi0}\{1+f\frac{m_{u}+m_{s}}{m}\},\\
f^{2}_{K^{0}}=f^{2}_{\pi0}\{1+f\frac{m_{d}+m_{s}}{m}\}, \\
f^{2}_{\eta_{8}}=f^{2}_{\pi0}\{1+{1\over3}f\frac{m_{u}+
m_{d}+4m_{s}}{m}\}
\end{eqnarray}

Substituting Eqs.(25,28,29,30)
to Eq.(18), to the second order in quark masses
the masses of pion and kaon are obtained
\begin{eqnarray}
\lefteqn{m^{2}_{\pi^{\pm}}={4\over f^{2}_{\pi0}}\{-{1\over3}
<\bar{\psi}\psi>(m_{u}+m_{d})-{F^{2}\over4}(m_{u}+m_{d})^{2}
+{f\over3}<\bar{\psi}\psi>{1\over m}(m_{u}+m_{d})^{2}
\},}\nonumber \\
&&m^{2}_{\pi^{0}}={4\over f^{2}_{\pi0}}\{-{1\over3}
<\bar{\psi}\psi>(m_{u}+m_{d})-{F^{2}\over2}(m^{2}_{u}+m^{2}_{d})
+{f\over3}<\bar{\psi}\psi>{1\over m}(m_{u}+m_{d})^{2}
\},\nonumber \\
&&m^{2}_{K^{+}}={4\over f^{2}_{\pi0}}\{-{1\over3}
<\bar{\psi}\psi>(m_{u}+m_{s})-{F^{2}\over4}(m_{u}+m_{s})^{2}
+{f\over3}<\bar{\psi}\psi>{1\over m}(m_{u}+m_{s})^{2}
\},\nonumber \\
&&m^{2}_{K^{0}}={4\over f^{2}_{\pi0}}\{-{1\over3}
<\bar{\psi}\psi>(m_{d}+m_{s})-{F^{2}\over4}(m_{d}+m_{s})^{2}
+{f\over3}<\bar{\psi}\psi>{1\over m}(m_{d}+m_{s})^{2}
\},\nonumber \\
&&m^{2}_{\eta_{8}}={4\over f^{2}_{\pi0}}\{-{1\over3}
<\bar{\psi}\psi>{1\over3}(m_{u}+
m_{d}+4m_{s})-{F^{2}\over4}{2\over3}(m^{2}_{u}+m^{2}_{d}
+4m_{s}^{2})\nonumber \\
&&+{f\over3}<\bar{\psi}\psi>{1\over9}{1\over m}
(m_{u}+m_{d}+4m_{s})^{2}\}\nonumber \\
&&={4\over f^{2}_{\pi0}}\{-{1\over3}
<\bar{\psi}\psi>{1\over3}(m_{u}+
m_{d}+4m_{s})
+({f\over m}{1\over3}<\bar{\psi}\psi>-{F^{2}\over 4})
{1\over9}
(m_{u}+m_{d}+4m_{s})^{2}\nonumber \\
&&-{2F^{2}\over9}[{1\over2}(m_{u}+m_{d})-m_{s}]^{2}
-{F^{2}\over12}(m_{d}-m_{u})^{2}\}.
\end{eqnarray}
The dependence of the meson masses(31) on the quark masses
are the same as the ones
presented in Ref.(1(b)).
The mass difference of charged pion and neutral pion is found from
Eqs.(31)
\begin{equation}
m^{2}_{\pi^{\pm}}-m^{2}_{\pi^{0}}=(1-{2c\over g})^{-1}(m_{d}
-m_{u})^{2}.
\end{equation}
Comparing with Eqs.(31) with the Eqs.(10.7,10.8) of Ref.(1(b)),
following
coefficients of the ChPT are predicted
\begin{eqnarray}
L_{4}=0,\;\;\;\;L_{6}=0,\\
L_{5}=\frac{f^{2}_{\pi0}f}{8mB_{0}},\\
L_{8}=-\frac{F^{2}}{16B_{0}^{2}},\\
3L_{7}+L_{8}=-\frac{F^{2}}{16B_{0}^{2}},\;\;\;L_{7}=0,
\end{eqnarray}
where
\begin{equation}
B_{0}={4\over f^{2}_{\pi0}}(-{1\over3})<\bar{\psi}\psi>.
\end{equation}
Using Eq.(37), $L_{5}$ and $L_{8}$ are written as
\begin{eqnarray}
L_{5}=\frac{1}{32Q}(1-{2c\over g})\{(1-{2c\over g})^{2}
(1-{1\over2\pi^{2}g^{2}})-(1-{2c\over g})+{4\over \pi^{2}}Q
(1-{c\over g})\},\\
L_{8}=-\frac{1}{1536g^{2}Q^{2}}(1-{2c\over g})^{2},
\end{eqnarray}
where
\begin{equation}
Q=-{1\over3}{m\over F^{4}}<\bar{\psi}\psi>.
\end{equation}
The Eqs.(19,20) show that Q is a function of the universal coupling
constant g only. The numerical value of Q is 4.54.

Both $L_{5}$ and $L_{8}$ are at $O(N_{C})$. $L_{4}$, $L_{6}$, and
$L_{7}$ are from loop diagrams of mesons and are at $O(1)$ in large
$N_{C}$ expansion.

By inputting the values of
$m^{2}_{\pi^{+}}$, $m^{2}_{K^{+}}$,
$m^{2}_{K^{0}}$, and $f^{2}_{\pi}$
(\(f_{\pi}=0.184GeV\)) into the Eqs.(18,25)
we are able to determine the values
of the three current quark masses and $F^{2}$. The universal coupling
constant \(g=0.39\) is chosen to fit the decay rate of $\rho\rightarrow
e^{-}e^{+}$. The results are
\begin{eqnarray}
F^{2}=0.0537GeV^{2},\\
m_{u}=0.91MeV,\;\;\;m_{d}=2.15MeV,\;\;\;m_{s}=52.2MeV.
\end{eqnarray}
The values of the quark masses are less than their traditional
values[20]. As pointed out in Ref.[20], the absolute values
of $m_{q}$ are model dependent. The values of $m_{q}$ determined
by lattice gauge calculation are smaller too[21].
Using Eqs.(41,42), the $f_{K}$ is calculated to be
\[f_{K}=0.215GeV,\]
\[f_{K}=1.17f_{\pi}.\]
The experimental data is \(f_{K}=1.22 f_{\pi}\). The deviation of the
theoretical result from the experiment is about
$4\%$.
To determine
$m_{\eta}$ and $f_{\eta}$
the mixing
between $\eta_{8}$ and $\eta_{0}$ is needed to be taken into account.
From the effective Lagrangian(17) the terms related to the masses of
$\eta$ and $\eta'$ are derived
\begin{eqnarray}
\lefteqn{{\cal L}_{\eta\eta'}=-{1\over2}\eta^{2}_{8}\{-{1\over3}
<\bar{\psi}\psi>{1\over3}(m_{u}+m_{d}+4m_{s})-{F^{2}\over6}
(m^{2}_{u}+m^{2}_{d}+4m^{2}_{s})\}}\nonumber \\
&&-{1\over2}\eta^{2}_{0}\{-{1\over3}
<\bar{\psi}\psi>{2\over3}(m_{u}+m_{d}+m_{s})-{F^{2}\over3}
(m^{2}_{u}+m^{2}_{d}+m^{2}_{s})+m^{4}_{g}\}\nonumber \\
&&-\eta_{0}\eta_{8}\{{\sqrt{2}\over9}
<\bar{\psi}\psi>(m_{u}+m_{d}-2m_{s})-{\sqrt{2}F^{2}\over12}
(m^{2}_{u}+m^{2}_{d}-2m^{2}_{s})\},
\end{eqnarray}
where $m_{g}$ is a parameter resulted by the $U(1)$ problem[22],
in Eq.(43) $\eta_{0}$
and $\eta_{8}$ are not normalized yet. The mixing is introduced as
\begin{equation}
\eta_{8}=\eta cos\theta-\eta' sin\theta,\;\;\;\eta_{0}=\eta
sin\theta+\eta' cos\theta.
\end{equation}
Eliminating the crossing term of $\eta\eta'$ and inputting
$m_{\eta'}$ to determine $m_{g}$, we obtain
\begin{eqnarray}
\lefteqn{m^{2}_{\eta}={4\over f^{2}_{\eta}}\{cos^{2}\theta[-{1\over9}
<\bar{\psi}\psi>(m_{u}+m_{d}+4m_{s})-{F^{2}\over6}(m^{2}_{u}+
m^{2}_{d}+4m^{2}_{s})]\}}\nonumber \\
&&+sin^{2}\theta[-{2\over9}<\bar{\psi}\psi>(m_{u}+m_{d}+m_{s})
-{F^{2}\over3}(m^{2}_{u}
+m^{2}_{d}+m^{2}_{s})+m^{4}_{g}]\nonumber \\
&&+{\sqrt{2}\over3}sin2\theta[{1\over3}<\bar{\psi}\psi>(m_{u}+m_{d}
-2m_{s})+{F^{2}\over4}(m^{2}_{u}+m^{2}_{d}-2m^{2}_{s})]\},
\nonumber \\
&&f^{2}_{\eta}=F^{2}(1-{2c\over g})\{1+[(1-{2c\over g})
(1-{1\over2\pi^{2}g^{2}})-1](sin^{2}\alpha{1\over m}(m_{u}+m_{d})
+cos^{2}\alpha{2\over m}m_{s})\}\nonumber \\
&&+{1\over \pi^{2}}(1-{2c\over g})
(1-{c\over g}){4\over3f^{2}_{\pi0}}\{cos^{2}
\theta(-{1\over3})<\bar{\psi}\psi>(m_{u}+m_{d}+4m_{s})\nonumber \\
&&+sin^{2}\theta
[-{1\over3}<\bar{\psi}\psi>2(m_{u}+m_{d}+m_{s})+3m^{4}_{g}] \nonumber \\
&&+{\sqrt{2}\over3}sin2\theta<\bar{\psi}\psi>(m_{u}+m_{d}-2m_{s})\},
\end{eqnarray}
where \(\alpha=\theta+35.26^{0}\)(\(sin35.26^{0}={1\over\sqrt{3}}\)).
The numerical results are
\begin{eqnarray}
\theta=7.84^{0},
m^{th}_{\eta}=0.569GeV,\;\;\;m^{exp}_{\eta}=0.548GeV,\\
f_{\eta}=0.24GeV=1.3f_{\pi}.
\end{eqnarray}
The theoretical value of $m_{\eta}$ deviates from the experiment by
$3.8\%$. The values of the mixing angle and $f_{\eta}$ can be tested
by the decay $\eta\rightarrow2\gamma$. Because the quark mass
correction has been taken into account in calculating $f_{\eta}$,
the effect of quark masses has to be included in calculating
the decay amplitude. This study is beyond the scope of this paper.

According to Refs.[1(b),3], $L_{9}$ and $L_{10}$ are determined by
$<r^{2}>_{\pi}$ and the amplitudes of pion radiative decay,
$\pi^{-}\rightarrow e^{-}\gamma\nu$
\begin{eqnarray}
L_{9}={f^{2}_{\pi}\over48}<r^{2}_{\pi}>,\\
L_{9}={1\over32\pi^{2}}{R\over F^{V}}, \\
L_{10}={1\over32\pi^{2}}{F^{A}\over F^{V}}-L_{9},
\end{eqnarray}
where R, $F^{V}$, and $F^{A}$
are $r_{A}$, $h_{V}$,and $h_{A}$ of Ref.[3] respectively.
In Ref.[19] we have derived the expression of pion radius as
\begin{equation}
<r^{2}>_{\pi}={6\over m^{2}_{\rho}}+{3\over\pi^{2}f^{2}_{\pi}}
\{(1-{2c\over g})^{2}-4\pi^{2}c^{2}\}
\end{equation}
which agrees with the data very well.
Using Eqs.(48,51), it is predicted[23]
\begin{equation}
L_{9}={1\over4}cg+{1\over16\pi^{2}}
\{(1-{2c\over g})^{2}-4\pi^{2}c^{2}\},
\end{equation}
On the other hand, $L_{9}$ is determined by
Eq.(49) too. The three form factors of the decay amplitude of
$\pi^{-}\rightarrow e^{-}\gamma\nu$ are presented in our paper[19]
\begin{eqnarray}
F^{V}=\frac{m_{\pi}}{2\sqrt{2}\pi^{2}f_{\pi}},\\
F^{A}={1\over2\sqrt{2}\pi^{2}}{m_{\pi}\over f_{\pi}}{m^{2}_{\rho}
\over m^{2}_{a}}(1-{2c\over g})(1-{1\over2\pi^{2}g^{2}})^{-1},
\\
R={g^{2}\over\sqrt{2}}{m_{\pi}\over f_{\pi}}{m^{2}_{\rho}\over
m^{2}_{a}}\{{2c\over g}+{1\over\pi^{2}g^{2}}(1-{2c\over g})\}
(1-{1\over2\pi^{2}g^{2}})^{-1}+\sqrt{2}cg{m_{\pi}\over f_{\pi}}
[24].
\end{eqnarray}
Using the mass formula of the $a_{1}$ meson(in the chiral limit)[10]
\[(1-{1\over2\pi^{2}g^{2}})m^{2}_{a}={F^{2}\over g^{2}}+m^{2}_
{\rho},\]
it is obtained
\begin{equation}
{m^{2}_{\rho}\over m^{2}_{a}}(1-{1\over2\pi^{2}g^{2}})^{-1}
=1-{2c\over g}.
\end{equation}
Substituting Eq.(56) into R(55), it is derived
\begin{equation}
R={1\over3\sqrt{2}}m_{\pi}f_{\pi}<r^{2}>_{\pi}.
\end{equation}
This is the expression obtained by applying PCAC to this process[25].
Therefore, the coefficient $L_{9}$ derived from Eq.(49) is the same
as the one obtained from Eq.(48).

Using Eq.(56), the ratio ${F^{A}\over F^{V}}$ is written as
\begin{equation}
{F^{A}\over F^{V}}=(1-{2c\over g})^{2}.
\end{equation}
The coefficient $L_{10}$ is found from Eq.(50)
\begin{equation}
L_{10}=-{1\over4}cg+{1\over4}c^{2}-{1\over32\pi^{2}}
(1-{2c\over g})^{2}.
\end{equation}
Both $L_{9}$ and $L_{10}$ are at $O(N_{C})$.

In the expressions of the coefficients(13,15,16,38,39,52,59)
there are two
parameters: g and $c$ which have been determined from
$\rho\rightarrow e^{-}e^{+}$ and the ratio
$\frac{f^{2}_{\pi}}{m^{2}_{\rho}}$ respectively.
The coefficients of the ChPT are completely determined by
this effective chiral theory. The numerical values of the coefficients
of ChPT are presented in table 3.
\begin{table}[h]
\begin{center}
\caption{Predictions of the Values of the coefficients}
\begin{tabular}{|c|c|c|c|c|c|c|c|c|c|}  \hline
$10^{3}L_{1}$&$10^{3}L_{2}$&$10^{3}L_{3}$
&$10^{3}L_{4}$&$10^{3}L_{5}$&$10^{3}L_{6}$
&$10^{3}L_{7}$&$10^{3}L_{8}$&$10^{3}L_{9}$
&$10^{3}L_{10}$  \\ \hline
3.0&1.4&-8.8&0&4.77&0&0&-0.079&8.3&-7.1\\ \hline
\end{tabular}
\end{center}
\end{table}

To summarize the results.
The orders of the coefficients of the ChPT
in large $N_{C}$ expansion are predicted as $L_{1,2,3,5,8,9,10}\sim
O(N_{C})$ and $L_{4,6}$ are $O(1)$. These predictions agree with
Ref.[1(b)]. It is also predicted that $L_{7}\sim O(1)$ in this paper.
The numerical value of $L_{7}$ determined in Ref.[1(b)](see Table 1)
is smaller. It is interesting to notice that \(L_{1}+{1\over2}L_{3}
=-1.4\) which is compatible with the value shown in Table 2.
It is necessary to point out that the expressions of the coefficients
$L_{4,5,6,7,8}$ are derived from Eq.(31), however, Eq.(18) is used to
fit the masses of the pseudoscalar mesons. It is found in this paper
that this way
leads to consistent determination of the quark masses.
There are no new parameters in
predicting the coefficients of the ChPT.
The quark masses determined in this paper are closer
to the lower bound obtained by lattice gauge calculation. Theoretical
values of $f_{K}$ and $m^{2}_{\eta}$ agree with data within about
$4\%$.

The author wishes to thank B.Holstein for discussion.
This research was partially
supported by DOE Grant No. DE-91ER75661.


\begin{thebibliography}{40}
\bibitem{} In Eq.(67) of Ref.[19] the factor ${1\over2}$ should be
taken off and there should be an
additional term from the form factor
\(\frac{m^{2}_{\rho}}{m^{2}_{\rho}-k^{2}}\).
\end{thebibliography}
\end{document}